\documentclass[aps,prl,twocolumn,amsmath,amssymb,superscriptaddress,citeautoscript,floatfix,nofootinbib,nobalancelastpage]{revtex4-2}
\usepackage{amsbsy}
\usepackage[dvipsnames]{xcolor}
\usepackage{wasysym}
\usepackage{bm}
\usepackage{textcomp}
\usepackage{tikz}
\usetikzlibrary{arrows.meta}
\usetikzlibrary{calc}
\usepackage{gensymb}
\usepackage[breaklinks,colorlinks = true,linkcolor = red,urlcolor=Blue,citecolor=Blue]{hyperref}

\usepackage{array}
\usepackage{graphicx}
\usepackage{subfigure}
\usepackage{float}
\usepackage{soul}
\usepackage{braket}

\newcommand\eq[1]{\begin{align}#1\end{align}}
\newcommand{\sr}[1]{{\color{black}{#1}}}

\newcommand{\jzz}{J}
\newcommand{\ed}{\varepsilon_\delta}
\newcommand{\ndcl}{N_{\rm cl}^\delta}
\newcommand{\nqcl}{N_{\rm cl}^{\cal Q}}
\newcommand{\qcl}{{\cal Q}_{\rm cl}}
\newcommand{\rhod}{\rho_\delta}
\newcommand{\rhoq}{\rho_{|\cal Q|}}

\begin{document}

\title{Arrested Relaxation  in a Disorder-Free Coulomb Spin Liquid}

\author{Souvik Kundu}
\email{souvik.kundu@icts.res.in}
\affiliation{International Centre for Theoretical Sciences, Tata Institute of Fundamental Research, Bengaluru 560089, India}

\author{Arnab Seth}
\email{aseth65@gatech.edu}
\affiliation{International Centre for Theoretical Sciences, Tata Institute of Fundamental Research, Bengaluru 560089, India}
\affiliation{School of Physics, Georgia Institute of Technology, Atlanta, Georgia 30332, USA}

\author{Sthitadhi Roy}
\email{sthitadhi.roy@icts.res.in}
\affiliation{International Centre for Theoretical Sciences, Tata Institute of Fundamental Research, Bengaluru 560089, India}
\affiliation{Max-Planck-Institut f\"{u}r Physik komplexer Systeme, N\"{o}thnitzer Stra{\ss}e 38, 01187 Dresden, Germany}

\author{Subhro Bhattacharjee}
\email{subhro@icts.res.in}
\affiliation{International Centre for Theoretical Sciences, Tata Institute of Fundamental Research, Bengaluru 560089, India}

\author{Roderich Moessner}
\email{moessner@pks.mpg.de}
\affiliation{Max-Planck-Institut f\"{u}r Physik komplexer Systeme, N\"{o}thnitzer Stra{\ss}e 38, 01187 Dresden, Germany}

\date{\today}

\begin{abstract}
    We investigate Coulomb spin liquids in classical spin-3/2 ice and show that the enlarged on-site Hilbert space gives rise to a qualitatively new class of such phases. Beyond the conventional magnetic monopoles of spin-1/2 ice, the system hosts additional low-energy crystal-field excitations, whose interplay with monopoles significantly modifies both equilibrium and non-equilibrium properties. Following a thermal quench, we find a pronounced dynamical arrest manifested in an exponentially long-lived {athermal} plateau in spin autocorrelations. This constitutes a rare example of dynamical arrest in a short-range interacting, disorder-free system. We demonstrate that the arrested dynamics originate from novel composite excitation structures unique to spin-3/2 ice and from kinetically constrained relaxation pathways that require activated processes. Our results establish higher-spin ice as a fertile platform for realizing unconventional Coulomb spin liquids and dynamical arrest without quenched disorder.

\end{abstract}

\maketitle

\paragraph*{Introduction:} 

The emergence of thermodynamic descriptions, such as those manifested in Gibbs ensembles, underpins much of the universal physics of interacting many-body systems in equilibrium~\cite{landau1980statistical}. 
Equally fundamental is the dynamics of the approach to such an equilibrium, as it allows for a richer classification of matter in terms of dynamical phases in a broad range of settings and contexts~\cite{hohenberg1977theory,hinrichsen2000nonequilibirum,henkel2008non,parameswaran2017eigenstate,moessner2017equilibration,parameswaran2018many}. In generic condensed matter, the process of equilibration is rapid. This naturally renders scenarios where the relaxation to equilibrium is ultraslow or arrested fundamentally interesting.

Over the last several decades, it has been comprehensively understood that quenched disorder can provide a robust mechanism for slow or arrested thermalisation -- archetypal examples include (spin) glasses and quantum localised systems (see Refs.~\cite{mezard1986spinglass,binder1986spin,berthier2011theoretical,berthier2011dynamical,charbonneau2017glass,garrahan2018aspects,lee1985disordered,nandkishore2015many,abanin2017recent,abanin2019colloquium,roy2024fock,sierant2025many} for reviews and references therein). The situation in (quenched) disorder-free systems is far less clear in this respect. While systems with kinetic constraints or Hilbert-space fragmentation provide interesting alternatives~\cite{ritort2003glassy,garrahan2011kinetically,lan2018quantum,pai2019localisation,sala2020ergodicity,roy2020strong,khemani2020localisation,pancotti2020quantum,scherg2021observing,deger2022arresting,biswas2022beyond}, these routes are almost always fine-tuned. This leaves open the possibility of a complex interplay of microscopic energy scales leading to emergent constrained dynamics of excitations, as a simple yet robust pathway to arrested relaxation to equilibrium. The presence of competing energy scales allows for non-trivial equilibrium correlations in the extensively degenerate ground state manifold and makes them natural candidates for slow relaxation by allowing for energetic barriers within the degenerate manifold without further fine-tuning. Therefore, frustrated spin-systems~\cite{diep2013frustrated} can provide simple and tunable models of disorder-free slow relaxation, offering controlled insights into the microscopic ingredients of glassy physics in the absence of quenched disorder~\cite{rau2016spin}.

A striking example of frustration-mediated, non-trivial correlation occurs in three-dimensional classical Coulomb spin liquids that are characterised by an emergent electromagnetism~\cite{PhysRevLett.91.167004,PhysRevB.68.184512,hermele2004}. This leads to anomalous low-temperature properties such as power-law spin-spin correlations~\cite{PhysRevB.71.014424,PhysRevLett.95.217201} and gapped {magnetic-monopole} excitations~\cite{castelnovo2008}. 
A paradigmatic realisation is classical spin ice (CSI), a
frustrated spin-$1/2$ Ising antiferromagnet on the pyrochlore lattice, which hosts a robust Coulomb phase and admits
several material realizations in rare earth magnets~\cite{gardner,rau2019,ho2ti2o7,dy2ti2o7,BramwellGingras2001}. 
From a broader perspective, however, the spin-$1/2$ nature of the local degrees of freedom is not essential to the emergence of Coulomb phases. 
This raises the fundamental question of whether higher-spin systems can host novel Coulomb spin liquids with properties absent in their spin-$1/2$ counterparts, driven solely by the enlarged local Hilbert space. 
Motivated also in part by
candidate materials, such as Tb$_2$Ti$_2$O$_7$, where
microscopic energetics~\cite{PhysRevB.76.184436,gardner99,PhysRevLett.99.237202,HoTbDy,tbtioprincep} result in two low-energy doublets that are separated by a small energy scale,
yielding an effective spin-3/2 manifold, we address this question by studying the thermodynamics and quench dynamics of classical spin-$3/2$ ice.

\begin{figure}
	\centering
	\includegraphics[width=\linewidth]{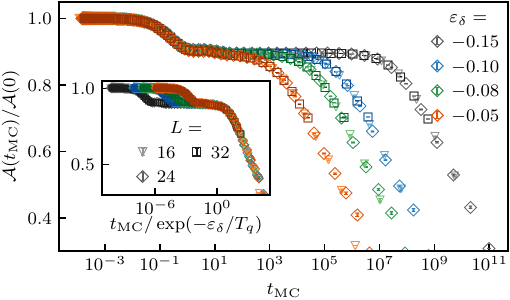}
	\caption{The spin autocorrelation, ${\cal A}(t_{\rm MC}) = \braket{S^z_i(t_{\rm MC})S^z_i(0)}$ (where $\braket{\cdot}$ denotes average over space and Monte-Carlo runs), shows dynamical arrest following a thermal quench from a high-temperature $T_i=1$ to a final temperature of $T_q=0.01$ (measured in units such that $J=1$). This is manifested in a long-lived plateau in Monte-Carlo time, $t_{\rm MC}$. Different colours denote different values of $\varepsilon_{\delta}$ (legend in the main panel) whereas different markers/intensities denote different system sizes $L$ (legend in inset). The height as well as the temporal extent of the plateau are independent of $L$. The scaling collapse in the inset shows that the plateau extends up to timescales which grow exponentially with $|\varepsilon_\delta|$.
    }
    \label{fig:autocorr}
\end{figure}

In this Letter, we uncover a new class of Coulomb spin liquids in classical spin-$3/2$ ice, enabled by the substantially richer low-energy excitation spectrum afforded by the enlarged on-site Hilbert space. 
A salient manifestation of this physics is the dynamical arrest following a thermal quench, characterised by an exponentially long-lived {athermal} plateau in the relaxation of spin autocorrelations (Fig.~\ref{fig:autocorr}). 
Remarkably, this realises dynamical arrest in a short-range interacting, disorder-free system where such behaviour is typically rather elusive. 
We show that this behaviour originates from novel excitation structures unique to spin-$3/2$ ice and their constrained relaxation pathways.
More specifically, the larger onsite Hilbert space leads to a new branch of low-energy excitations distinct from the monopoles, which we refer to as the $\delta$ excitations, whose relaxation dynamics are, in fact, strongly correlated with that of the monopoles. 
In particular, they form composite excitation motifs which can only mutually annihilate.
However, their dynamics are activated and follow an Arrhenius law, which in turn leads to an athermal plateau with a lifetime which is exponentially large in the Arrhenius energy barrier.

\paragraph*{Model:} As a concrete setting, we consider the classical $S=3/2$ Ising antiferromagnet on the pyrochlore lattice, described by the Hamiltonian,
\begin{align}
    H=\jzz\sum_{\langle i,j\rangle}S^z_iS^z_j+\frac{\Delta}{2}\sum_{i}\left((S^z_i)^2-\frac{1}{4}\right)\,,
    \label{eq:classham}
\end{align}
where $\jzz>0$ denotes the antiferromagnetic exchange between spins quantised along the local $[111]$ directions, 
and $\Delta$ controls the on-site splitting between the ${S^z=\pm 1/2}$ and ${S^z=\pm 3/2}$ doublets. 
The interplay of $\Delta$ and $\jzz$ is made transparent by rewriting Eq.~\ref{eq:classham} as
\eq{
H=\sum_{\boxtimes}\left[\frac{\jzz}{2}\left(\mathcal{Q}_\boxtimes\right)^2+\frac{\varepsilon_\delta}{4}\sum_{i\in\boxtimes}\left(\left(S^z_i\right)^2-\frac{1}{4}\right)\right]\,,
}
up to a constant, where $\mathcal{Q}_\boxtimes=\eta_{\boxtimes}\sum_{i\in\boxtimes}S^z_i=-6,\ldots,+6$ denotes the magnetic monopole charge on tetrahedron $\boxtimes$, with $\eta_\boxtimes=+1~(-1)$ for $A~(B)$ tetrahedra~\cite{castelnovo2012}. 

\paragraph{Low-energy excitations:} 
The presence of these two doublets and the interplay of the energy scales, $J$ and $\Delta$, lead to the $\delta$ excitations, which are a branch of low-energy local crystal field excitations. Unlike the monopoles, these are neutral under the emergent electromagnetism and cost an energy $\propto |\ed|$ with $\ed\equiv \Delta-2\jzz$.
Below, we show how the coupled quench dynamics of monopoles and $\delta$ excitations underpin the dynamical arrest.

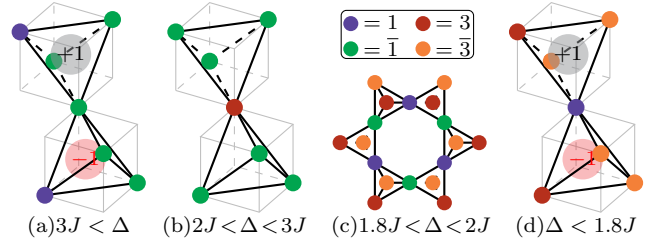
\begin{figure}
\begin{tikzpicture}[scale=0.9]
\coordinate (A000) at (0,0);
\coordinate (A100) at (-0.5,-0.433);
\coordinate (A110) at (0.36602540378443876, -0.6830127018922192);
\coordinate (A010) at (0.8660254037844387, -0.25);
\coordinate (A001) at (0.0, 0.8660254037844387);
\coordinate (A101) at (-0.49999999999999994, 0.4330127018922194);
\coordinate (A111) at (0.36602540378443876, 0.18301270189221952);
\coordinate (A011) at (0.8660254037844387, 0.6160254037844388);
\coordinate (B000) at (-0.36602540378443876, 1.549038105676658);
\coordinate (B100) at (-0.8660254037844387, 1.1160254037844386);
\coordinate (B110) at (0.0, 0.8660254037844387);
\coordinate (B010) at (0.5, 1.299038105676658);
\coordinate (B001) at (-0.36602540378443876, 2.4150635094610964);
\coordinate (B101) at (-0.8660254037844387, 1.9820508075688774);
\coordinate (B111) at (0.0, 1.7320508075688774);
\coordinate (B011) at (0.49999999999999994, 2.165063509461097);

\draw[gray!50] (A100)--(A110)--(A010);
\draw[gray!50,dashed] (A010)--(A000)--(A100);
\draw[gray!50] (A101)--(A111)--(A011)--(A001)--cycle;
\draw[gray!50] (A010)--(A011);
\draw[gray!50] (A110)--(A111);
\draw[gray!50] (A100)--(A101);
\draw[gray!50,dashed] (A000)--(A001);

\draw[gray!50] (B100)--(B110)--(B010);
\draw[gray!50,dashed] (B010)--(B000)--(B100);
\draw[gray!50] (B101)--(B111)--(B011)--(B001)--cycle;
\draw[gray!50] (B010)--(B011);
\draw[gray!50] (B110)--(B111);
\draw[gray!50] (B100)--(B101);
\draw[gray!50,dashed] (B000)--(B001);

\fill[Red,fill opacity=0.3] (0.1,0.1) circle[radius=0.3];
\draw[black,thick] (A100)--(A010)--(A111)--cycle;
\draw[black,thick] (A100)--(A001);
\draw[black,thick] (A010)--(A001);
\draw[black,thick] (A111)--(A001);
\draw[black,thick] (B110)--(B101)--(B011)--cycle;
\draw[black,thick,dashed] (B000)--(B110);
\draw[black,thick,dashed] (B000)--(B011);
\draw[black,thick,dashed] (B000)--(B101);

\fill[Violet] (A100) circle[radius=0.125];
\fill[Green] (A010) circle[radius=0.125];
\fill[Green] (A111) circle[radius=0.125];
\fill[Green] (A001) circle[radius=0.125];

\fill[Green] (B000) circle[radius=0.125];
\fill[Green] (B011) circle[radius=0.125];
\fill[Violet] (B101) circle[radius=0.125];
\fill[Gray,fill opacity=0.5] (-0.26602540378443876+0.15, 1.649038105676658) circle[radius=0.3];
\node at (-0.26602540378443876+.15, 1.649038105676658) {\footnotesize{$+1$}};
\node at (.1,.1) {\footnotesize{{\color{red}$-1$}}};
\node at (0,-.9) {\footnotesize{(a)$3J<\Delta$}};
\end{tikzpicture}
\begin{tikzpicture}[scale=0.9]
\coordinate (A000) at (0,0);
\coordinate (A100) at (-0.5,-0.433);
\coordinate (A110) at (0.36602540378443876, -0.6830127018922192);
\coordinate (A010) at (0.8660254037844387, -0.25);
\coordinate (A001) at (0.0, 0.8660254037844387);
\coordinate (A101) at (-0.49999999999999994, 0.4330127018922194);
\coordinate (A111) at (0.36602540378443876, 0.18301270189221952);
\coordinate (A011) at (0.8660254037844387, 0.6160254037844388);
\coordinate (B000) at (-0.36602540378443876, 1.549038105676658);
\coordinate (B100) at (-0.8660254037844387, 1.1160254037844386);
\coordinate (B110) at (0.0, 0.8660254037844387);
\coordinate (B010) at (0.5, 1.299038105676658);
\coordinate (B001) at (-0.36602540378443876, 2.4150635094610964);
\coordinate (B101) at (-0.8660254037844387, 1.9820508075688774);
\coordinate (B111) at (0.0, 1.7320508075688774);
\coordinate (B011) at (0.49999999999999994, 2.165063509461097);

\draw[gray!50] (A100)--(A110)--(A010);
\draw[gray!50,dashed] (A010)--(A000)--(A100);
\draw[gray!50] (A101)--(A111)--(A011)--(A001)--cycle;
\draw[gray!50] (A010)--(A011);
\draw[gray!50] (A110)--(A111);
\draw[gray!50] (A100)--(A101);
\draw[gray!50,dashed] (A000)--(A001);

\draw[gray!50] (B100)--(B110)--(B010);
\draw[gray!50,dashed] (B010)--(B000)--(B100);
\draw[gray!50] (B101)--(B111)--(B011)--(B001)--cycle;
\draw[gray!50] (B010)--(B011);
\draw[gray!50] (B110)--(B111);
\draw[gray!50] (B100)--(B101);
\draw[gray!50,dashed] (B000)--(B001);

\draw[black,thick] (A100)--(A010)--(A111)--cycle;
\draw[black,thick] (A100)--(A001);
\draw[black,thick] (A010)--(A001);
\draw[black,thick] (A111)--(A001);
\draw[black,thick] (B110)--(B101)--(B011)--cycle;
\draw[black,thick,dashed] (B000)--(B110);
\draw[black,thick,dashed] (B000)--(B011);
\draw[black,thick,dashed] (B000)--(B101);

\fill[Green] (A100) circle[radius=0.125];
\fill[Green] (A010) circle[radius=0.125];
\fill[Green] (A111) circle[radius=0.125];
\fill[BrickRed] (A001) circle[radius=0.125];
\fill[Green] (B000) circle[radius=0.125];
\fill[Green] (B011) circle[radius=0.125];
\fill[Green] (B101) circle[radius=0.125];
\node at (0,-.9) {\footnotesize{(b)$2J\!<\!\Delta\!<\!3 J$}};
\end{tikzpicture}
\begin{tikzpicture}[scale=1.6]
\draw[black,thick](0,0)--(0,1)--(0.866,0.5)--cycle;
\draw[black,thick](-0.577/2,0.5)--(0.577,1)--(0.577,0.)--cycle;

\draw[black,thick](0.1666/1.732,5/6)--(0,2/3);
\draw[black,thick](0.1666/1.732,5/6)--(0.5/1.732,5/6);
\draw[black,thick](0.1666/1.732,5/6)--(0,1);
\fill[BrickRed] (0.1666/1.732,5/6) circle[radius=0.125/2];
\fill[Orange] (0,1) circle[radius=0.125/2];

\draw[black,thick,dashed](2.5/3/1.732,5/6)--(0.5/1.732,5/6);
\draw[black,thick,dashed](2.5/3/1.732,5/6)--(1/1.732,4/6);
\draw[black,thick,dashed](2.5/3/1.732,5/6)--(1/1.732,1);
\fill[BrickRed] (2.5/3/1.732,5/6) circle[radius=0.125/2];
\fill[Orange] (1/1.732,1) circle[radius=0.125/2];

\draw[black,thick](7/6/1.732,1.5/3)--(1/1.732,1/3);
\draw[black,thick](7/6/1.732,1.5/3)--(1.732/2,.5);
\draw[black,thick](7/6/1.732,1.5/3)--(1/1.732,2/3);
\fill[Orange] (7/6/1.732,1.5/3) circle[radius=0.125/2];
\fill[BrickRed] (1.732/2,.5) circle[radius=0.125/2];

\draw[black,thick,dashed](2.5/3/1.732,1/6)--(0.5/1.732,1/6);
\draw[black,thick,dashed](2.5/3/1.732,1/6)--(1/1.732,2/6);
\draw[black,thick,dashed](2.5/3/1.732,1/6)--(1/1.732,0);
\fill[Orange] (2.5/3/1.732,1/6) circle[radius=0.125/2];
\fill[BrickRed] (1/1.732,0) circle[radius=0.125/2];

\draw[black,thick](0.1666/1.732,1/6)--(0,1/3);
\draw[black,thick](0.1666/1.732,1/6)--(0.5/1.732,1/6);
\draw[black,thick](0.1666/1.732,1/6)--(0,0);
\fill[Orange] (0.1666/1.732,1/6) circle[radius=0.125/2];
\fill[BrickRed] (0,0) circle[radius=0.125/2];

\draw[black,thick,dashed](-1/6/1.732,1/2)--(0,1/3);
\draw[black,thick,dashed](-1/6/1.732,1/2)--(-0.5/1.732,1/2);
\draw[black,thick,dashed](-1/6/1.732,1/2)--(0,2/3);
\fill[Orange] (-1/6/1.732,1/2) circle[radius=0.125/2];
\fill[BrickRed] (-0.5/1.732,1/2) circle[radius=0.125/2];

\fill[Violet] (0,1/3) circle[radius=0.125/2];
\fill[Violet] (1/2/1.732,5/6) circle[radius=0.125/2];
\fill[Violet] (1/1.732,1/3) circle[radius=0.125/2];
\fill[Green] (0,2/3) circle[radius=0.125/2];
\fill[Green] (1/1.732,4/6) circle[radius=0.125/2];
\fill[Green] (1/2/1.732,1/6) circle[radius=0.125/2];

\fill[Violet] (-0.2,1.49) circle[radius=0.125/2];
\node at (0.05,1.5) {\footnotesize{$=1$}};
\fill[Green] (-0.2,1.28) circle[radius=0.125/2];
\node at (0.05,1.3) {\footnotesize{$=\overline{1}$}};

\fill[BrickRed] (0.4,1.49) circle[radius=0.125/2];
\node at (0.65,1.5) {\footnotesize{$=3$}};
\fill[Orange] (0.4,1.28) circle[radius=0.125/2];
\node at (0.65,1.3) {\footnotesize{$=\overline{3}$}};

\draw[rounded corners=2pt] (-0.28,1.2) rectangle (0.82,1.6); 
\node at (0.3,-.2) {\footnotesize{(c)$1.8J\!<\!\Delta\!<\!2J$}};
\end{tikzpicture}
\begin{tikzpicture}[scale=0.9]
\coordinate (A000) at (0,0);
\coordinate (A100) at (-0.5,-0.433);
\coordinate (A110) at (0.36602540378443876, -0.6830127018922192);
\coordinate (A010) at (0.8660254037844387, -0.25);
\coordinate (A001) at (0.0, 0.8660254037844387);
\coordinate (A101) at (-0.49999999999999994, 0.4330127018922194);
\coordinate (A111) at (0.36602540378443876, 0.18301270189221952);
\coordinate (A011) at (0.8660254037844387, 0.6160254037844388);
\coordinate (B000) at (-0.36602540378443876, 1.549038105676658);
\coordinate (B100) at (-0.8660254037844387, 1.1160254037844386);
\coordinate (B110) at (0.0, 0.8660254037844387);
\coordinate (B010) at (0.5, 1.299038105676658);
\coordinate (B001) at (-0.36602540378443876, 2.4150635094610964);
\coordinate (B101) at (-0.8660254037844387, 1.9820508075688774);
\coordinate (B111) at (0.0, 1.7320508075688774);
\coordinate (B011) at (0.49999999999999994, 2.165063509461097);

\draw[gray!50] (A100)--(A110)--(A010);
\draw[gray!50,dashed] (A010)--(A000)--(A100);
\draw[gray!50] (A101)--(A111)--(A011)--(A001)--cycle;
\draw[gray!50] (A010)--(A011);
\draw[gray!50] (A110)--(A111);
\draw[gray!50] (A100)--(A101);
\draw[gray!50,dashed] (A000)--(A001);

\draw[gray!50] (B100)--(B110)--(B010);
\draw[gray!50,dashed] (B010)--(B000)--(B100);
\draw[gray!50] (B101)--(B111)--(B011)--(B001)--cycle;
\draw[gray!50] (B010)--(B011);
\draw[gray!50] (B110)--(B111);
\draw[gray!50] (B100)--(B101);
\draw[gray!50,dashed] (B000)--(B001);

\fill[Red,fill opacity=0.3] (0.1,0.1) circle[radius=0.3];

\draw[black,thick] (A100)--(A010)--(A111)--cycle;
\draw[black,thick] (A100)--(A001);
\draw[black,thick] (A010)--(A001);
\draw[black,thick] (A111)--(A001);
\draw[black,thick] (B110)--(B101)--(B011)--cycle;
\draw[black,thick,dashed] (B000)--(B110);
\draw[black,thick,dashed] (B000)--(B011);
\draw[black,thick,dashed] (B000)--(B101);

\fill[BrickRed] (A100) circle[radius=0.125];
\fill[Orange] (A010) circle[radius=0.125];
\fill[Orange] (A111) circle[radius=0.125];
\fill[Violet] (A001) circle[radius=0.125];
\fill[Orange] (B000) circle[radius=0.125];
\fill[Orange] (B011) circle[radius=0.125];
\fill[BrickRed] (B101) circle[radius=0.125];
\fill[Gray,fill opacity=0.5] (-0.26602540378443876+.15, 1.649038105676658) circle[radius=0.3];
\node at (-0.26602540378443876+.15, 1.649038105676658) {\footnotesize{$+1$}};
\node at (.1,.1) {\footnotesize{{\color{red}$-1$}}};
\node at (0,-.9) {\footnotesize{(d)$\Delta < 1.8 J$}};
\end{tikzpicture}
\caption{Elementary lowest-energy excitations in different parameter regimes. The circles of different colours denote the four different states of a local $S=3/2$ degree of freedom as indicated in the legend. The translucent grey and red circles in (a) and (d) denote a magnetic monopole charge of ${\cal Q}_\boxtimes=1$ and $-1$ respectively in the corresponding tetrahedra. For ${1.8J<\Delta<2J}$ (panels (b) and (c)), the lowest-energy excitations are charge-neutral $\delta$ excitations as exemplified by the absence of any monopole.}
\label{fig:excitation}
\end{figure}

At $\varepsilon_\delta=0$, the ground state (GS) on each tetrahedron is a 44-fold degenerate {\it extended ice manifold}, comprising all charge-neutral configurations of the form $\ket{1\bar{1} 1\bar{1}}$, $\ket{3\bar{3} 3\bar{3}}$, $\ket{1\bar{1} 3\bar{3}}$, $\ket{3\bar{1} \bar{1}\bar{1}}$, and their partners related by symmetry and permutations, where we use ${1(\bar{1})\equiv \tfrac{1}{2}(\tfrac{-1}{2})}$ and ${3(\bar{3})\equiv \tfrac{3}{2}(\tfrac{-3}{2})}$.
The lowest-energy excitations above this manifold are magnetic monopoles with ${\cal Q}_{\boxtimes}=\pm 1$, which cost energy $\propto \jzz$ and are 80-fold degenerate per tetrahedron.
For $\varepsilon_\delta\neq0$, the $44$ fold degeneracy of the extended ice manifold is partially lifted, and the GS manifold is that of the usual CSI but formed of the $S^z=\pm\tfrac{1}{2}$ doublets for $\varepsilon_\delta>0$ and $S^z=\pm\tfrac{3}{2}$ for $\varepsilon_\delta<0$. 
For sufficiently small $|\varepsilon_\delta|$, the remaining configurations of the extended ice manifold survive as the lowest-energy charge-neutral excitations, which we identify as the $\delta$ excitations. 
Their structure, however, depends qualitatively on the sign of $\varepsilon_\delta$, as summarised in Fig.~\ref{fig:excitation}.

For $\ed>0$, the elementary $\delta$ excitations on top of the $\ket{1\bar{1}1\bar{1}}$ manifold are composed of configurations of the kind $\ket{3\bar{1}\bar{1}\bar{1}}$ and $\ket{\bar{3}111}$ (Fig.~\ref{fig:excitation}(b)), and they form the lowest-energy excitations for ${2J<\Delta<3J}$. For $\Delta>3J$, the usual spin-1/2 CSI physics is recovered; magnetic monopoles with ${\cal Q}_\boxtimes=\pm 1$ constitute the lowest energy excitations (Fig.~\ref{fig:excitation}(a)). Note that the elementary monopoles do not involve the $3(\bar{3})$ states and are therefore devoid of any $\delta$ excitation.

The low-energy excitation structure for $\ed<0$ is substantially richer. 
Here, the ground-state manifold consists of $\ket{3\bar{3}3\bar{3}}$ configurations, and the low-energy, charge-neutral excitations necessarily involve at least a pair of $\delta$ excitations on each tetrahedron.
As a consequence, isolated $\delta$ excitations are forbidden, and charge-neutral excitations can only appear as clusters of $\delta$ excitations.
The lowest-energy excitation of this kind on the pyrochlore lattice is a closed loop of length 6, with an energy cost of $6|\ed|$, consisting of alternating $1$ and $\bar{1}$ spins [Fig.~\ref{fig:excitation}(c)].

A key feature of the $\ed<0$ regime is that ${\cal Q}_{\boxtimes}=\pm 1$ monopoles are confined.
Pairs of ${\cal Q}_{\boxtimes}=\pm1$ monopoles must be connected by a string of $\delta$ excitations, leading to an energy cost proportional to the string length. 
The minimal such excitation is therefore a pair of ${\cal Q}_{\boxtimes}=\pm 1$ monopoles on adjacent tetrahedra with a single $\delta$ excitation on the site shared by them (Fig.~\ref{fig:excitation}(d)), which costs an energy $J+|\ed|$. 
For $|\varepsilon_\delta|<\jzz/5$ (equivalently $1.8\jzz<\Delta<2\jzz$), the charge-neutral 6-$\delta$ excitation loop (energy cost of $6|\ed|$) is therefore energetically favoured over the monopole pairs with a single $\delta$ excitation.
Given this rich structure of the low-energy excitations for ${\ed<0}$, we will focus on this regime for the quench dynamics and return to the case of ${\ed>0}$ briefly later. In the rest of the paper we work in units such that $J=1$.

\paragraph{Excitations at high temperature:} Indeed, as we discuss now, the constrained relaxation dynamics of these low-energy excitations underpin the {arrested} dynamics. 
However, to obtain insights into the thermal quench dynamics, it is equally important to understand the statistics of the high-energy excitations at the initial high temperature prior to the quench~\cite{em}. 
Note that for $\ed<0$, monopole pairs are necessarily accompanied by $\delta$ excitations except for the energetically unfavourable ${\cal Q}_\boxtimes=\pm3,6$ pairs.
We therefore organise the $\delta$ excitations in a finite-temperature configuration into clusters: a cluster is defined as the contiguous set of all tetrahedra that are connected by at least a $\delta$ excitation on the shared sites. 
We will denote the number of $\delta$ and monopole excitations in a cluster as $\ndcl$ and $\nqcl$ respectively, and the total monopole charge in the cluster as $\qcl$.
In addition, $n(\ndcl,\nqcl)$ denotes the number of clusters with $\ndcl$ $\delta$ excitations and $\nqcl$ monopoles, and $n_{\rm tot}$ as the total number of clusters.
We define $P(\ndcl,\nqcl)\equiv n(\ndcl,\nqcl)/n_{\rm tot}$ as the joint probability of finding clusters characterised by $\ndcl$ and $\nqcl$. 
Similarly, $P(\qcl,\nqcl)$ is the joint probability of finding clusters with $\nqcl$ monopoles with total charge $\qcl$.

\begin{figure}
\includegraphics[width=\linewidth]{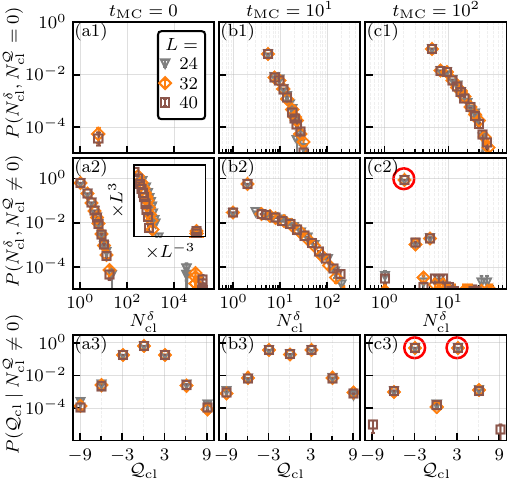}
\caption{The joint probability $P(\ndcl,\nqcl)$ of the cluster sizes and their monopole content as well the conditional probability $P(\qcl|\nqcl\neq 0)$ of the monopole charge at three different Monte-Carlo times (different columns) following a thermal quench from $T_i=1$ to $T_q=0.01$ for $\ed=-0.08$. The $t_{\rm MC}=0$ data (shown in panels (a)) corresponds to equilibrium at $T_i$. The inset in (a2) evinces the system-spanning clusters with $\ndcl\sim O(L^3)$. The middle and right columns correspond to representative $t_{\rm MC}$ during the initial fast quench and metastable plateau respectively. The red circles in (c2) and (c3) highlight clusters with $(\ndcl,\nqcl,\qcl)=(2,3,\pm 3)$ as discussed in the text.
}
\label{fig:cluster-dist}
\end{figure}

The results for these distributions in equilibrium at a high temperature of ${T=1}$, which is also our initial temperature for the thermal quench, and $\ed=-0.08$ are shown in Fig.~\ref{fig:cluster-dist}(a1)-(a3).
Decomposing $P(\ndcl)=P(\ndcl,\nqcl=0)+P(\ndcl,\nqcl\neq 0)$,
we find that the former constitutes a tiny fraction of the clusters and is dominated by $\ndcl=6$ (Fig.~\ref{fig:cluster-dist}(a1)); these are, in fact, the monopole-free $6$-$\delta$ loop excitations illustrated in Fig.~\ref{fig:excitation}(b).
Barring these, almost all clusters contain monopoles, and there is a finite density of small clusters ($\ndcl\sim O(1)$) with an exponential size distribution, together
with an $O(1)$ number of system-spanning clusters with $\ndcl\sim O(L^3)$; the volume scaling of the latter is confirmed in the inset in Fig.~\ref{fig:cluster-dist}(a2).
The conditional distribution, $P(\qcl|\nqcl\neq 0)$, in Fig.~\ref{fig:cluster-dist}(a3) shows that, within the monopole containing clusters, $\qcl$ necessarily takes values that are integer multiples of $3$.
This is due to the fact that clusters containing $\qcl\neq 3k$ with $k\in \mathbb{Z}$ are necessarily connected to another cluster with charge $3k-\qcl$ via a string of $\delta$ excitations; as such, they belong to a single larger cluster with a total charge $3k$. 
On the other hand, there is no such string necessary for the $\qcl=\pm3k$ pair, as they are deconfined.

\begin{figure}
\includegraphics[width=\linewidth]{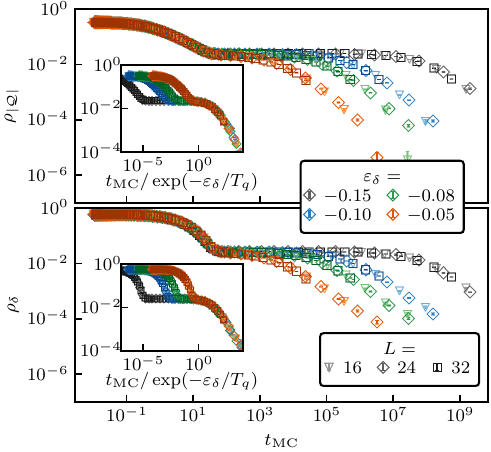}
\caption{{Arrested relaxation} dynamics of the excitations under thermal quench from $T_i=1$ to $T_q=0.01$ for $\ed<0$, manifested in the exponentially long lived,  {athermal	} plateaux in the densities of $\delta$ and monopole excitation, $\rhod$ and $\rhoq$, respectively. Different colours denote different values of $\ed$ and different markers/intensities denote different system sizes. The insets show the lifetime of the plateaux is $\tau = \exp(|\ed|/T_q)$, identical to that of the spin-autocorrelation in Fig.~\ref{fig:autocorr}, as scaling $t_{\rm MC}$ by $\tau$ collapses the knees of the plateaux.
}
\label{fig:rho-del-Q-dyn}
\end{figure}

\paragraph{Thermal quench:}
To understand the relaxation dynamics of the excitations following a thermal quench from an initial temperature $T_i=1$ to a final low temperature $T_q=0.01$, we study the dynamics of the densities of the $\delta$ and monopole excitations, $\rhod$ and $\rhoq$, as a function of MC time $t_{\rm MC}$, using the waiting-time Monte Carlo algorithm~\cite{Dall_2001} with single spin-flip updates~\cite{supp}. 
The results, presented in Fig.~\ref{fig:rho-del-Q-dyn}, show the same {arrested} dynamics as the spin-autocorrelation, ${\cal A}(t_{\rm MC})$ (Fig.~\ref{fig:autocorr}) manifested in a system-size independent, {athermal} plateau in $\rhod$ and $\rhoq$ as a function of $t_{\rm MC}$.
In fact, the temporal extent of the plateau, exponentially large as $\exp(|\ed|/T_q)$, is also the same as that in ${\cal A}(t_{\rm MC})$.
This is evident as scaling the time axis as $t_{\rm MC}/e^{|\ed|/T_q}$ collapses the temporal endpoints of the plateaux (knees of the plateaux) and the subsequent decay of the excitations onto a single curve for different values of $\ed$, as shown in the insets in Fig.~\ref{fig:rho-del-Q-dyn}.
This suggests that the dynamical arrest in the spin autocorrelation is underpinned by the same in the relaxation of the excitations. 
Insights into the latter are obtained by studying the evolution of the probability distributions $P(\ndcl,\nqcl)$ and $P(\qcl|\nqcl)$ with MC time as follows.

\sr{In the early-time regime, both $\rho_\delta$ and $\rho_{\cal Q}$ decay rapidly with Monte Carlo time $t_{\rm MC}$. 
During this stage, large system-spanning clusters fragment into smaller ones, as evidenced by the disappearance of the peak at $\ndcl\sim\mathcal{O}(L^3)$ in Fig.~\ref{fig:cluster-dist}(b1). 
A small fraction of these fragments are monopole free; once formed, such clusters are dynamically inert and therefore persist throughout the arrested regime, leading to $P(\ndcl,\nqcl=0)$ remaining essentially unchanged across the plateau [Fig.~\ref{fig:cluster-dist}(c1)].

The remaining clusters contain magnetic monopoles and continue to evolve via two distinct processes. 
First, tetrahedra carrying large monopole charges, $|{\cal Q}_{\boxtimes}|=3,6$, reorganise into configurations involving $|{\cal Q}_{\boxtimes}|=1$ monopoles connected by strings of $\delta$ excitations. 
Second, elementary monopoles within these clusters undergo pair annihilation, reducing the tension associated with the $\delta$ strings. 
As a result, by the end of the fast-quench regime the dynamics becomes dominated by metastable motifs characterised by $(\ndcl,\nqcl,\qcl)=(2,3,\pm3)$. 
These motifs account for the prominent features of the cluster distributions on the plateau, highlighted by the red circles in Fig.~\ref{fig:cluster-dist}(c2)–(c3).

The subsequent relaxation is controlled by the slow annihilation of these metastable clusters, as well as of the monopole-free ones. 
The key process enabling this is the collective hopping of the $(\ndcl,\nqcl,\qcl)=(2,3,\pm3)$ motifs, which allows motifs of opposite total charge to encounter and annihilate [Fig.~\ref{fig:Q3-hop-annihilate}]. 
As they diffuse, these motifs can also absorb monopole-free clusters [Fig.~\ref{fig:Q3-hop-Q0annihilate}]. 
Crucially, however, the hopping of a $(2,3,\pm3)$ motif is an activated process: its motion requires the transient creation of an additional $\delta$ excitation, costing an energy $|\varepsilon_\delta|$, illustrated as,}
\begin{center}
\begin{tikzpicture}[scale=0.7]
\def\s{3}

\coordinate (A) at (0,0);
\coordinate (B) at (1,0);
\coordinate (C) at (0.5,-1.732/2);
\coordinate (D) at (0.5,-1.732/6);
\coordinate (E) at (0,-1.732);
\coordinate (F) at (1,-1.732);
\coordinate (G) at (0.5,-1.732*5/6);
\coordinate (b) at (-1,0);
\coordinate (c) at (-0.5,1.732/2);
\coordinate (d) at (-0.5,1.732/6);
\coordinate (e) at (0,1.732);
\coordinate (f) at (-1,1.732);
\coordinate (g) at (-0.5,1.732*5/6);

\fill[Red!10] (f)--(e)--(A)--(b)--cycle;
\fill[Red!10] (B)--(c)--(b)--(C)--cycle;
\fill[Gray,fill opacity=0.75] (g) circle[radius=0.3];
\fill[Gray,fill opacity=0.75] (d) circle[radius=0.3];
\fill[Gray,fill opacity=0.75] (D) circle[radius=0.3];
\draw[thick] (A)--(B)--(C)--cycle;
\draw[thick] (A)--(D)--(B)--(D)--(C);
\draw[thick] (E)--(F)--(C)--cycle;
\draw[Gray,thin] (C)--(G)--(E)--(G)--(F);
\draw[thick] (A)--(b)--(c)--cycle;
\draw[Gray,thin] (A)--(d)--(b)--(d)--(c);
\draw[thick] (e)--(f)--(c)--cycle;
\draw[thick] (c)--(g)--(e)--(g)--(f);
\fill[BrickRed] (e) circle[radius=0.125];
\fill[Orange] (f) circle[radius=0.125];
\fill[BrickRed] (g) circle[radius=0.125];
\fill[Green] (c) circle[radius=0.125];
\fill[Orange] (d) circle[radius=0.125];
\fill[BrickRed] (b) circle[radius=0.125];
\fill[Green] (A) circle[radius=0.125];
\fill[BrickRed] (B) circle[radius=0.125];
\fill[Orange] (D) circle[radius=0.125];
\fill[BrickRed] (C) circle[radius=0.125];
\fill[Orange] (E) circle[radius=0.125];
\fill[Orange] (G) circle[radius=0.125];
\fill[BrickRed] (F) circle[radius=0.125];

\coordinate (A) at (0+\s,0);
\coordinate (B) at (1+\s,0);
\coordinate (C) at (0.5+\s,-1.732/2);
\coordinate (D) at (0.5+\s,-1.732/6);
\coordinate (E) at (0+\s,-1.732);
\coordinate (F) at (1+\s,-1.732);
\coordinate (G) at (0.5+\s,-1.732*5/6);
\coordinate (b) at (-1+\s,0);
\coordinate (c) at (-0.5+\s,1.732/2);
\coordinate (d) at (-0.5+\s,1.732/6);
\coordinate (e) at (0+\s,1.732);
\coordinate (f) at (-1+\s,1.732);
\coordinate (g) at (-0.5+\s,1.732*5/6);
\fill[Gray,fill opacity=0.75] (g) circle[radius=0.3];
\fill[Gray,fill opacity=0.75] (d) circle[radius=0.3];
\fill[Gray,fill opacity=0.75] (G) circle[radius=0.3];
\draw[thick] (A)--(B)--(C)--cycle;
\draw[thick] (A)--(D)--(B)--(D)--(C);
\draw[thick] (E)--(F)--(C)--cycle;
\draw[Gray,thin] (C)--(G)--(E)--(G)--(F);
\draw[thick] (A)--(b)--(c)--cycle;
\draw[Gray,thin] (A)--(d)--(b)--(d)--(c);
\draw[thick] (e)--(f)--(c)--cycle;
\draw[thick] (c)--(g)--(e)--(g)--(f);
\fill[BrickRed] (e) circle[radius=0.125];
\fill[Orange] (f) circle[radius=0.125];
\fill[BrickRed] (g) circle[radius=0.125];
\fill[Green] (c) circle[radius=0.125];
\fill[Orange] (d) circle[radius=0.125];
\fill[BrickRed] (b) circle[radius=0.125];
\fill[Green] (A) circle[radius=0.125];
\fill[BrickRed] (B) circle[radius=0.125];
\fill[Orange] (D) circle[radius=0.125];
\fill[Violet] (C) circle[radius=0.125];
\fill[Orange] (E) circle[radius=0.125];
\fill[Orange] (G) circle[radius=0.125];
\fill[BrickRed] (F) circle[radius=0.125];

\coordinate (A) at (0+2*\s,0);
\coordinate (B) at (1+2*\s,0);
\coordinate (C) at (0.5+2*\s,-1.732/2);
\coordinate (D) at (0.5+2*\s,-1.732/6);
\coordinate (E) at (0+2*\s,-1.732);
\coordinate (F) at (1+2*\s,-1.732);
\coordinate (G) at (0.5+2*\s,-1.732*5/6);
\coordinate (b) at (-1+2*\s,0);
\coordinate (c) at (-0.5+2*\s,1.732/2);
\coordinate (d) at (-0.5+2*\s,1.732/6);
\coordinate (e) at (0+2*\s,1.732);
\coordinate (f) at (-1+2*\s,1.732);
\coordinate (g) at (-0.5+2*\s,1.732*5/6);
\fill[Gray,fill opacity=0.75] (g) circle[radius=0.3];
\fill[Gray,fill opacity=0.75] (D) circle[radius=0.3];
\fill[Gray,fill opacity=0.75] (G) circle[radius=0.3];
\draw[thick] (A)--(B)--(C)--cycle;
\draw[thick] (A)--(D)--(B)--(D)--(C);
\draw[thick] (E)--(F)--(C)--cycle;
\draw[Gray,thin] (C)--(G)--(E)--(G)--(F);
\draw[thick] (A)--(b)--(c)--cycle;
\draw[Gray,thin] (A)--(d)--(b)--(d)--(c);
\draw[thick] (e)--(f)--(c)--cycle;
\draw[thick] (c)--(g)--(e)--(g)--(f);
\fill[BrickRed] (e) circle[radius=0.125];
\fill[Orange] (f) circle[radius=0.125];
\fill[BrickRed] (g) circle[radius=0.125];
\fill[Green] (c) circle[radius=0.125];
\fill[Orange] (d) circle[radius=0.125];
\fill[BrickRed] (b) circle[radius=0.125];
\fill[Violet] (A) circle[radius=0.125];
\fill[BrickRed] (B) circle[radius=0.125];
\fill[Orange] (D) circle[radius=0.125];
\fill[Violet] (C) circle[radius=0.125];
\fill[Orange] (E) circle[radius=0.125];
\fill[Orange] (G) circle[radius=0.125];
\fill[BrickRed] (F) circle[radius=0.125];

\coordinate (A) at (0+3*\s,0);
\coordinate (B) at (1+3*\s,0);
\coordinate (C) at (0.5+3*\s,-1.732/2);
\coordinate (D) at (0.5+3*\s,-1.732/6);
\coordinate (E) at (0+3*\s,-1.732);
\coordinate (F) at (1+3*\s,-1.732);
\coordinate (G) at (0.5+3*\s,-1.732*5/6);
\coordinate (b) at (-1+3*\s,0);
\coordinate (c) at (-0.5+3*\s,1.732/2);
\coordinate (d) at (-0.5+3*\s,1.732/6);
\coordinate (e) at (0+3*\s,1.732);
\coordinate (f) at (-1+3*\s,1.732);
\coordinate (g) at (-0.5+3*\s,1.732*5/6);
\fill[Red!10] (F)--(E)--(A)--(B)--cycle;
\fill[Red!10] (B)--(c)--(b)--(C)--cycle;
\fill[Gray,fill opacity=0.75] (d) circle[radius=0.3];
\fill[Gray,fill opacity=0.75] (D) circle[radius=0.3];
\fill[Gray,fill opacity=0.75] (G) circle[radius=0.3];
\draw[thick] (A)--(B)--(C)--cycle;
\draw[thick] (A)--(D)--(B)--(D)--(C);
\draw[thick] (E)--(F)--(C)--cycle;
\draw[Gray,thin] (C)--(G)--(E)--(G)--(F);
\draw[thick] (A)--(b)--(c)--cycle;
\draw[Gray,thin] (A)--(d)--(b)--(d)--(c);
\draw[thick] (e)--(f)--(c)--cycle;
\draw[thick] (c)--(g)--(e)--(g)--(f);
\fill[BrickRed] (e) circle[radius=0.125];
\fill[Orange] (f) circle[radius=0.125];
\fill[BrickRed] (g) circle[radius=0.125];
\fill[Orange] (c) circle[radius=0.125];
\fill[Orange] (d) circle[radius=0.125];
\fill[BrickRed] (b) circle[radius=0.125];
\fill[Violet] (A) circle[radius=0.125];
\fill[BrickRed] (B) circle[radius=0.125];
\fill[Orange] (D) circle[radius=0.125];
\fill[Violet] (C) circle[radius=0.125];
\fill[Orange] (E) circle[radius=0.125];
\fill[Orange] (G) circle[radius=0.125];
\fill[BrickRed] (F) circle[radius=0.125];

\draw[-Stealth, thick, Red] (0.5+1,-1.732/2) -- (0.5+\s-1,-1.732/2);
\node at (0.5+\s/2,-1.732/2+0.3) {\footnotesize{activated}};
\node at (0.5+\s/2,-1.732/2-0.3) {\footnotesize{hopping}};
\node at (0.5+\s/2,-1.732/2-0.7) {\footnotesize{$\Delta E\!=\!|\ed|$}};
\draw[-Stealth, thick] (0.5+\s+1,-1.732/2) -- (0.5+2*\s-1,-1.732/2);
\draw[-Stealth, thick] (0.5+2*\s+1,-1.732/2) -- (0.5+3*\s-1,-1.732/2);

\end{tikzpicture}
\end{center}
(where the colour coding matches Fig.~\ref{fig:excitation}). 
This activation barrier leads to a characteristic timescale $\tau\sim\exp(|\varepsilon_\delta|/T_q)$ at the quench temperature $T_q$, which sets the lifetime of the {athermal} plateau observed in the arrested dynamics and the diffusion time scale in the post-plateau relaxation dynamics and the subsequent decay of the plateau.

\begin{figure}
\includegraphics[width=\linewidth]{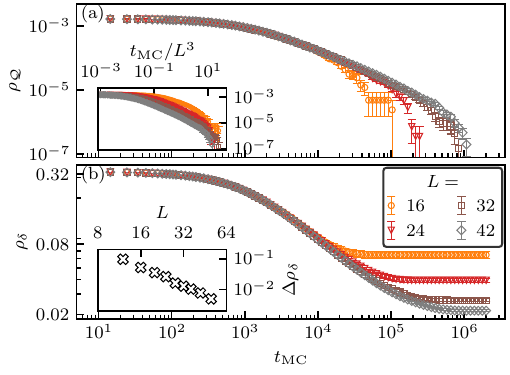}
\caption{Dynamics of (a) $\rhoq$ and (b) $\rhod$ as function of $t_{\rm MC}$ following a thermal quench from $T_i=0.08$ to $T_q=0.005$, for $\ed = 0.012$. The inset in (a) shows the final crash of $\rhoq$ for different $L$ collapses when $t_{\rm MC}$ is rescaled by $t_R\sim O(L^3)$. The inset in (b) shows the residual plateau in $\rhod$, quantified by $\Delta\rho_\delta = \rhod(t_{\rm MC}\to\infty)-\rho_{\rm eq}$ vanishes as $L\to\infty$ polynomially.}
\label{fig:ed>0-qd}
\end{figure}

\paragraph{Absence of dynamical arrest for ${\ed\!\!>\!0}$:}
The dynamical arrest observed for $\varepsilon_\delta<0$ relies crucially on the fact that isolated $\delta$ excitations and elementary magnetic monopoles are forbidden, which tightly couples their relaxation dynamics. 
This constraint is absent for $\varepsilon_\delta>0$: both isolated $\delta$ excitations and deconfined ${\cal Q}_{\boxtimes}=\pm1$ monopoles are allowed, with the latter undergoing diffusion and pair annihilation. 
As a consequence, monopoles in a quenched system relax rapidly.

While an isolated $\delta$ excitation can decay only via the creation of a neighbouring monopole–antimonopole pair, a process that requires a finite activation energy, the initial thermal ensemble already contains a finite density of monopoles. 
In the early-time regime following a quench, the dominant decay channel for $\delta$ excitations is therefore provided by diffusing monopoles passing through them, as illustrated in Fig.~\ref{fig:ed>0-del-relax}. 
Crucially, in the thermodynamic limit this finite initial monopole density is sufficient to eliminate all $\delta$ excitations, as demonstrated by the finite-size scaling of the quench dynamics shown in Fig.~\ref{fig:ed>0-qd}.

For a finite system, the monopole density $\rho_{\cal Q}$ relaxes completely on a characteristic timescale $t_R\sim L^3$ [Fig.~\ref{fig:ed>0-qd}(a)]. 
For $t_{\rm MC}\ll t_R$, the decay of monopoles is governed by a symmetric diffusion--annihilation process, closely analogous to that in classical nearest-neighbour spin-$1/2$ ice~\cite{castelnovo2010quench}. 
During this stage, the $\delta$-excitation density $\rho_\delta$ decays concomitantly, until all monopoles have annihilated. 
For $t_{\rm MC}>t_R$, finite-size simulations exhibit a residual plateau in $\rho_\delta$ above the equilibrium value $\rho_{\rm eq}$ at $T_q$; however, this difference from $\rho_{\rm eq}$ vanishes as $L\to\infty$, as shown by the finite-size scaling in the inset of Fig.~\ref{fig:ed>0-qd}(b).

We emphasise that these results are obtained for quenches from an initial temperature satisfying $\varepsilon_\delta\ll T_i\ll\jzz$, for which the initial monopole density is exponentially suppressed relative to $\rho_\delta$ by a factor $\sim\exp(-\jzz/T_i)$. 
Nevertheless, the diffusion of this dilute monopole population suffices to fully relax the $\delta$ excitations in the thermodynamic limit. 
The upshot is that for $\varepsilon_\delta>0$, the relaxation of $\delta$ excitations is controlled entirely by the diffusive motion of pre-existing monopoles, thereby precluding any dynamical arrest.

\paragraph{Discussion:} 
In summary, we have shown an explicit example of a short-range disorder-free spin model with frustrated spin interactions and a materials-based motivation that exhibits dynamical arrest following a thermal quench. It provides a simple, and yet robust, instance of emergent kinetically constrained relaxation of low-energy excitations.

Our results indicate that the slow relaxation for $\ed<0$ is connected to the presence of excitation clusters with an Arrhenius energy barrier.
The size distribution of these clusters (see Fig.~\ref{fig:cluster-dist}) indicate the presence of spatial heterogeneity, statistics of which evolve in time. Given the presence of multiple timescales, corresponding to the evolution of these clusters, it is natural to ask if the present setting can be used as a minimal model to gain insights into the physics of dynamical heterogeneity.

From the point of view of exotic phases, our model realises different flavours of Coulomb liquids. In a contemporaneous manuscript~\cite{pandey2026Z3flux}, it was found that the model can host two topologically distinct Coulomb liquids at zero temperature but with no distinguishing signatures between them in conventional probes such as structure factors. Here, their strikingly different non-equilibrium dynamics can be used as a qualitatively distinguishing feature, also possibly experimentally. Relatedly, it will be interesting to study how the arrested dynamics affects the emergence of characteristic equilibrium properties, following a quench into the different Coulomb liquids.
Finally, the fate of the slow dynamics under the addition of coherent quantum (tunnelling)  processes, presents an obvious avenue for future studies.

\acknowledgements
We thank L. Jaubert, K. Damle, B. Gaulin and S. Nakatsuji for useful discussions and acknowledge funding from Max Planck Partner group grants between ICTS-TIFR, Bengaluru and MPIPKS, Dresden. S.K. acknowledges the computing resources provided by the Department of Theoretical Physics (DTP) at the Tata Institute of Fundamental Research and thanks J. Pandey and K. Damle for collaboration on a closely related work. S.R. acknowledges support from SERB-DST, Government of India, under Grant No. SRG/2023/000858. S.B. acknowledges support from Swarna Jayanti fellowship grant of SERB-DST (India) Grant No. SB/SJF/2021-22/12; DST, Government of India (Nano mission), under Project No. DST/NM/TUE/QM-10/2019 (C)/7. S. K., A.S., S.R. and S.B. acknowledge the support of the Department of Atomic Energy, Government of India, under project nos. RTI4019 and RTI4013. This work was in part supported by the Deutsche Forschungsgemeinschaft under grants SFB 1143 (project-id 247310070) and the cluster of excellence ctd.qmat (EXC 2147, project-id 390858490). 


\bibliography{ref}

\begin{thebibliography}{60}
\bibitem [{em()}]{em}%
  \BibitemOpen
  \href@noop {} {}\bibinfo {note} {Additional details pertaining to the
  thermodynamic properties of the model are presented in the End
  Matter.}\BibitemShut {Stop}%
\end{thebibliography}


\begin{center}
{\bf END MATTER}
\end{center}

The End Matter consists of two sections; in the first section, we present results for the thermodynamic properties of the spin-3/2 ice and in the second section, we show illustrative diagrams for the (activated) diffusion process which underpins the dynamical arrest.

\subsection{Thermodynamics \label{sec:eqbm-thermo}}

\begin{figure}
\includegraphics[width=\linewidth]{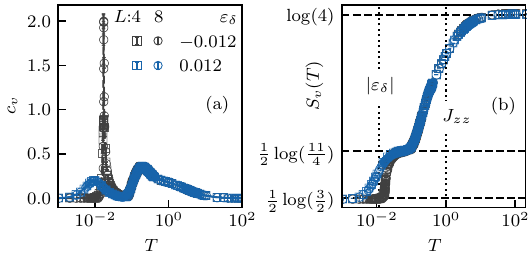}
     \caption{(a) Specific heat per site, $c_v$, plotted as a function of temperature $T$ for different system sizes and for $\varepsilon_{\delta}=\pm 0.012$. In both cases, $c_v$ exhibits two well-separated peaks, one at $T\sim|\varepsilon_{\delta}|$ and another at $T\sim \jzz$.  (b) Entropy density (defined in Eq.~\ref{eq:entropy}) as a function of $T$, showing a two-step crossover from the high-temperature paramagnetic regime to the conventional spin-1/2 ice phase via an intermediate spin-$3/2$ ice regime. In both panels, blue (black) curves correspond to $\varepsilon_{\delta}=0.012$ ($\varepsilon_{\delta}=-0.012$), and different symbols denote different system sizes.}
     \label{fig:cv_entropy}
\end{figure}

In this section we present details of the low-temperature thermodynamics of the Hamiltonian in Eq.~\ref{eq:classham}, obtained from extensive Monte Carlo simulations~\cite{supp}. As discussed in the main text, for $|\varepsilon_\delta|\ll \jzz$, the system exhibits distinct low-energy excitation structures for $\varepsilon_\delta>0$ and $\varepsilon_\delta<0$. These differences lead to qualitatively distinct temperature-driven, multistage crossover behaviour in the two regimes.
This is captured by the specific heat (per site),
\begin{eqnarray}
c_v = \frac{\langle E^2\rangle - \langle E\rangle^2}{N_s T^2},
\label{eq:cv}
\end{eqnarray}
and the thermal entropy density,
\begin{eqnarray}
S_v(T)=\log(4)-\int_T^{\infty}\frac{c_v}{T}\,dT,
\label{eq:entropy}
\end{eqnarray}
both of which are shown in Fig.~\ref{fig:cv_entropy} as functions of temperature $T$ for $\jzz=1.0$ and $\varepsilon_{\delta}=\pm 0.012$. In Eq.~\ref{eq:cv}, $\langle E\rangle$ is the average energy of the system and $N_s=4L^3$ is the total number of spins in the system.

In both regimes, the specific heat exhibits two well-separated peaks at temperatures $T_{\delta}\propto|\varepsilon_{\delta}|$ and $T_{\mathcal Q}\propto \jzz$. Correspondingly, the thermal entropy density displays two low-temperature plateaux at $S_v=\tfrac{1}{2}\log(3/2)$ and $S_v=\tfrac{1}{2}\log(11/4)$, associated with the conventional classical spin-1/2 ice state and the classical spin-$3/2$ ice state, respectively.

For $\varepsilon_\delta>0$, the crossover between the spin-$1/2$ ice plateau and the spin-$3/2$ ice plateau is smooth, characterized by a broad, system-size--independent peak in $c_v$. In contrast, for $\varepsilon_\delta<0$, the corresponding peak in $c_v$ is sharp and grows with system size, signalling a genuine thermodynamic phase transition within the extended ice manifold. This observation is consistent with very recent work~\cite{pandey2026Z3flux}, which demonstrated a first-order phase transition driven by the fugacity of $\delta$ excitations between two Coulomb phases with distinct topological properties within the $44$-vertex ice manifold.

The upshot of this is that, although the low-temperature Coulomb phases realized for $\varepsilon_\delta>0$ and $\varepsilon_\delta<0$ share the same thermal entropy density characteristic of the conventional CSI state, they nevertheless exhibit distinct topological characteristics; the emergence of dynamical arrest in the $\varepsilon_\delta<0$ regime underpins this topological distinction and may serve as an indirect probe for identifying the corresponding Coulomb phase.

\subsection{Activated-diffusion processes}

\begin{figure}[!t]
\input{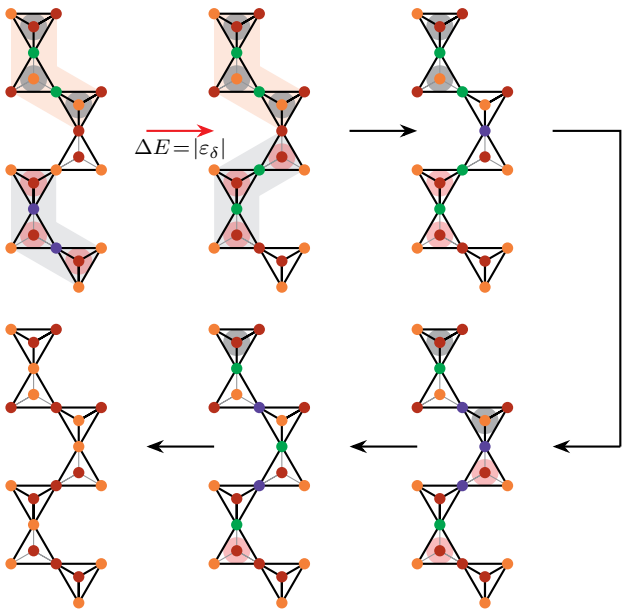}
\caption{Illustration of how activated hopping of the $(\ndcl,\nqcl,\qcl)=(2,3,\pm 3)$ clusters can lead to them pairwise annihilating each other with opposite total monopole charges}
\label{fig:Q3-hop-annihilate}
\end{figure}

In this section, we provide illustrative schematics of the different dynamical processes governing the post-plateau relaxation dynamics for $\varepsilon_\delta<0$. As discussed in the main text, this relaxation regime is controlled by the activated hopping of $(N^{\delta}_{\rm cl},N^{\mathcal{Q}}_{\rm cl},\mathcal{Q}_{\rm cl}) =(2,3,\pm 3)$ motifs. Such activated hopping leads to diffusion and subsequent annihilation of oppositely charged motifs, as illustrated in Fig.~\ref{fig:Q3-hop-annihilate}.

\begin{figure}
\input{activated_diffusion}
\caption{Illustration of how activated hopping of the $(\ndcl,\nqcl,\qcl)=(2,3,\pm 3)$ clusters can annihilate monopole-free clusters as the former diffuses.}
\label{fig:Q3-hop-Q0annihilate}
\end{figure}

We emphasise that this constitutes a slow-diffusion process when compared to the diffusion–annihilation dynamics associated with the relaxation of monopole excitations following a thermal quench in the usual nearest-neighbour (NN) CSI. The characteristic decay timescale of these motifs scales as $L^3 \exp(|\varepsilon_\delta|/T_q)$, reflecting the activation energy cost $|\varepsilon_\delta|$ associated with each unit hopping process. As a consequence, the effective diffusion constant is suppressed by the same exponential factor relative to that in the corresponding thermal quench of NN-CSI studied in Ref.~\cite{castelnovo2010quench}.

In addition to mutual annihilation, these motifs also participate in the elimination of monopole-free excitation clusters that remain dynamically inert after the initial fast-quench regime. This process proceeds in two stages. First, an activated hop allows a $(2,3,\pm 3)$ motif to attach to a monopole-free cluster $(N^{\delta}_{\rm cl},0,0)$, forming a larger composite cluster $(N^{\delta}_{\rm cl}+2,3,\pm 3)$. Once this composite cluster is formed, the monopoles undergo rapid directed motion within the cluster on an $\mathcal{O}(1)$ timescale to reduce the associated string tension, analogous to the shrinking of monopole-containing clusters observed during the initial fast-quench dynamics. This mechanism is schematically illustrated in Fig.~\ref{fig:Q3-hop-Q0annihilate}.

\begin{figure}
\begin{tikzpicture}[scale=0.7]
\def\s{3}

\coordinate (A) at (0,0);
\coordinate (B) at (1,0);
\coordinate (C) at (0.5,-1.732/2);
\coordinate (D) at (0.5,-1.732/6);
\coordinate (E) at (0,-1.732);
\coordinate (F) at (1,-1.732);
\coordinate (G) at (0.5,-1.732*5/6);
\coordinate (b) at (-1,0);
\coordinate (c) at (-0.5,1.732/2);
\coordinate (d) at (-0.5,1.732/6);
\coordinate (e) at (0,1.732);
\coordinate (f) at (-1,1.732);
\coordinate (g) at (-0.5,1.732*5/6);

\fill[Gray,fill opacity=0.75] (g) circle[radius=0.3];
\draw[thick] (A)--(B)--(C)--cycle;
\draw[thick] (A)--(D)--(B)--(D)--(C);
\draw[thick] (E)--(F)--(C)--cycle;
\draw[Gray,thin] (C)--(G)--(E)--(G)--(F);
\draw[thick] (A)--(b)--(c)--cycle;
\draw[Gray,thin] (A)--(d)--(b)--(d)--(c);
\draw[thick] (e)--(f)--(c)--cycle;
\draw[thick] (c)--(g)--(e)--(g)--(f);
\fill[Violet] (e) circle[radius=0.125];
\fill[Violet] (f) circle[radius=0.125];
\fill[Green] (g) circle[radius=0.125];
\fill[Violet] (c) circle[radius=0.125];
\fill[Violet] (d) circle[radius=0.125];
\fill[Violet] (b) circle[radius=0.125];
\fill[Orange] (A) circle[radius=0.125];
\fill[Violet] (B) circle[radius=0.125];
\fill[Green] (D) circle[radius=0.125];
\fill[BrickRed] (C) circle[radius=0.125];
\fill[Green] (E) circle[radius=0.125];
\fill[Green] (G) circle[radius=0.125];
\fill[Green] (F) circle[radius=0.125];

\coordinate (A) at (0+\s,0);
\coordinate (B) at (1+\s,0);
\coordinate (C) at (0.5+\s,-1.732/2);
\coordinate (D) at (0.5+\s,-1.732/6);
\coordinate (E) at (0+\s,-1.732);
\coordinate (F) at (1+\s,-1.732);
\coordinate (G) at (0.5+\s,-1.732*5/6);
\coordinate (b) at (-1+\s,0);
\coordinate (c) at (-0.5+\s,1.732/2);
\coordinate (d) at (-0.5+\s,1.732/6);
\coordinate (e) at (0+\s,1.732);
\coordinate (f) at (-1+\s,1.732);
\coordinate (g) at (-0.5+\s,1.732*5/6);

\fill[Gray,fill opacity=0.75] (d) circle[radius=0.3];
\draw[thick] (A)--(B)--(C)--cycle;
\draw[thick] (A)--(D)--(B)--(D)--(C);
\draw[thick] (E)--(F)--(C)--cycle;
\draw[Gray,thin] (C)--(G)--(E)--(G)--(F);
\draw[thick] (A)--(b)--(c)--cycle;
\draw[Gray,thin] (A)--(d)--(b)--(d)--(c);
\draw[thick] (e)--(f)--(c)--cycle;
\draw[thick] (c)--(g)--(e)--(g)--(f);
\fill[Violet] (e) circle[radius=0.125];
\fill[Violet] (f) circle[radius=0.125];
\fill[Green] (g) circle[radius=0.125];
\fill[Green] (c) circle[radius=0.125];
\fill[Violet] (d) circle[radius=0.125];
\fill[Violet] (b) circle[radius=0.125];
\fill[Orange] (A) circle[radius=0.125];
\fill[Violet] (B) circle[radius=0.125];
\fill[Green] (D) circle[radius=0.125];
\fill[BrickRed] (C) circle[radius=0.125];
\fill[Green] (E) circle[radius=0.125];
\fill[Green] (G) circle[radius=0.125];
\fill[Green] (F) circle[radius=0.125];

\coordinate (A) at (0+2*\s,0);
\coordinate (B) at (1+2*\s,0);
\coordinate (C) at (0.5+2*\s,-1.732/2);
\coordinate (D) at (0.5+2*\s,-1.732/6);
\coordinate (E) at (0+2*\s,-1.732);
\coordinate (F) at (1+2*\s,-1.732);
\coordinate (G) at (0.5+2*\s,-1.732*5/6);
\coordinate (b) at (-1+2*\s,0);
\coordinate (c) at (-0.5+2*\s,1.732/2);
\coordinate (d) at (-0.5+2*\s,1.732/6);
\coordinate (e) at (0+2*\s,1.732);
\coordinate (f) at (-1+2*\s,1.732);
\coordinate (g) at (-0.5+2*\s,1.732*5/6);

\fill[Gray,fill opacity=0.75] (D) circle[radius=0.3];
\draw[thick] (A)--(B)--(C)--cycle;
\draw[thick] (A)--(D)--(B)--(D)--(C);
\draw[thick] (E)--(F)--(C)--cycle;
\draw[Gray,thin] (C)--(G)--(E)--(G)--(F);
\draw[thick] (A)--(b)--(c)--cycle;
\draw[Gray,thin] (A)--(d)--(b)--(d)--(c);
\draw[thick] (e)--(f)--(c)--cycle;
\draw[thick] (c)--(g)--(e)--(g)--(f);
\fill[Violet] (e) circle[radius=0.125];
\fill[Violet] (f) circle[radius=0.125];
\fill[Green] (g) circle[radius=0.125];
\fill[Green] (c) circle[radius=0.125];
\fill[Violet] (d) circle[radius=0.125];
\fill[Violet] (b) circle[radius=0.125];
\fill[Green] (A) circle[radius=0.125];
\fill[Violet] (B) circle[radius=0.125];
\fill[Green] (D) circle[radius=0.125];
\fill[BrickRed] (C) circle[radius=0.125];
\fill[Green] (E) circle[radius=0.125];
\fill[Green] (G) circle[radius=0.125];
\fill[Green] (F) circle[radius=0.125];

\coordinate (A) at (0+3*\s,0);
\coordinate (B) at (1+3*\s,0);
\coordinate (C) at (0.5+3*\s,-1.732/2);
\coordinate (D) at (0.5+3*\s,-1.732/6);
\coordinate (E) at (0+3*\s,-1.732);
\coordinate (F) at (1+3*\s,-1.732);
\coordinate (G) at (0.5+3*\s,-1.732*5/6);
\coordinate (b) at (-1+3*\s,0);
\coordinate (c) at (-0.5+3*\s,1.732/2);
\coordinate (d) at (-0.5+3*\s,1.732/6);
\coordinate (e) at (0+3*\s,1.732);
\coordinate (f) at (-1+3*\s,1.732);
\coordinate (g) at (-0.5+3*\s,1.732*5/6);

\fill[Gray,fill opacity=0.75] (G) circle[radius=0.3];
\draw[thick] (A)--(B)--(C)--cycle;
\draw[thick] (A)--(D)--(B)--(D)--(C);
\draw[thick] (E)--(F)--(C)--cycle;
\draw[Gray,thin] (C)--(G)--(E)--(G)--(F);
\draw[thick] (A)--(b)--(c)--cycle;
\draw[Gray,thin] (A)--(d)--(b)--(d)--(c);
\draw[thick] (e)--(f)--(c)--cycle;
\draw[thick] (c)--(g)--(e)--(g)--(f);
\fill[Violet] (e) circle[radius=0.125];
\fill[Violet] (f) circle[radius=0.125];
\fill[Green] (g) circle[radius=0.125];
\fill[Green] (c) circle[radius=0.125];
\fill[Violet] (d) circle[radius=0.125];
\fill[Violet] (b) circle[radius=0.125];
\fill[Green] (A) circle[radius=0.125];
\fill[Violet] (B) circle[radius=0.125];
\fill[Green] (D) circle[radius=0.125];
\fill[Violet] (C) circle[radius=0.125];
\fill[Green] (E) circle[radius=0.125];
\fill[Green] (G) circle[radius=0.125];
\fill[Green] (F) circle[radius=0.125];

\draw[-Stealth, thick] (0.5+1,-1.732/2) -- (0.5+\s-1,-1.732/2);
\draw[-Stealth, thick] (0.5+\s+1,-1.732/2) -- (0.5+2*\s-1,-1.732/2);
\draw[-Stealth, thick] (0.5+2*\s+1,-1.732/2) -- (0.5+3*\s-1,-1.732/2);

\end{tikzpicture}
\caption{Illustration of how the diffusion of ${\cal Q}_{\boxtimes}=\pm 1$ monopoles can annihilate $\delta$ excitations in the $\ed>0$ regime.}
\label{fig:ed>0-del-relax}
\end{figure}
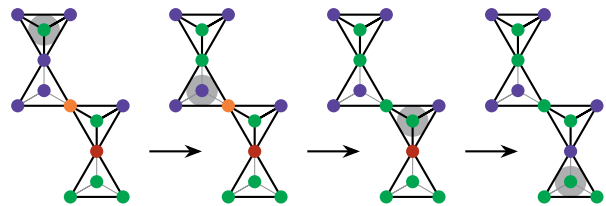

Finally, note that for $\ed>0$, the $\delta$ excitations are annihilated as the deconfined, freely diffusing monopoles pass through it, as illustrated in Fig.~\ref{fig:ed>0-del-relax}.

\clearpage\newpage

\setcounter{equation}{0}
\setcounter{figure}{0}
\setcounter{page}{1}
\renewcommand{\theequation}{S\arabic{equation}}
\renewcommand{\thefigure}{S\arabic{figure}}
\renewcommand{\thesection}{S\arabic{section}}
\renewcommand{\thepage}{S\arabic{page}}

\onecolumngrid
    \begin{center}
        {{\bf Supplementary Material: Arrested Relaxation  in a Disorder-Free Coulomb Spin Liquid}}
        \smallskip
        
        Souvik Kundu, Arnab Seth, Sthitadhi Roy, Subhro Bhattacharjee, and Roderich Moessner
    \end{center}

\twocolumngrid

This Supplementary Material, containing two sections, briefly outlines the numerical methods used in this work. 
Section I describes the classical Monte Carlo simulations used to obtain equilibrium thermodynamic properties, while Section II details the single-spin-flip dynamics employed to study the relaxation dynamics following thermal quenches.

\section{I. Numerical methods for Thermodynamics \label{sec:numerics}}

We perform Monte Carlo simulations using a combination of local Metropolis spin updates and non-local worm updates defined on the dual diamond lattice. In a local update, a pyrochlore site is chosen at random and its spin is changed to another allowed value with Metropolis acceptance probability. This update efficiently equilibrates the system at high temperatures, where a finite density of monopoles is present. At low temperatures, however, the acceptance rate of local updates becomes strongly suppressed, necessitating the use of non-local worm updates \cite{Sandvik_Moessner_2006,Alet_Ikhlef_etal_2006,Rakala_Damle_2017,Morita_Lee_Damle_Kawashima_2023,Kundu_Damle_2025,Kundu_Damle_2025_arxiv}.

Since each site carries an $S=3/2$ degree of freedom, there are multiple ways to change the spin configuration at a given site, each defining a distinct species of worm characterised by the charge carried at its ends. Specifically, changing the spin on a site by an amount $\partial S$ ($\partial S=\pm1,\pm2,\pm3$) creates a pair of charges $\pm\partial S$ on the two tetrahedra adjacent to that site. We refer to these two charges as the worm tail and the worm head. Worm propagation corresponds to a weighted random walk of one of these charges (the worm head), implemented via a sequence of spin flips that satisfy local detailed balance. The worm moves along the links of the dual diamond lattice, with both the worm head and tail residing on dual diamond sites.

We define the construction and directed motion of each worm species in a simple manner that explicitly preserves detailed balance. Worm construction begins by choosing a random tetrahedron (i.e., a site of the dual diamond lattice). If the chosen tetrahedron initially carries no charge, a worm of species $\pm\partial S$ is initiated by flipping one of the flippable spins associated with that tetrahedron. This spin flip creates a pair of charges $\pm\partial S$: one charge resides on the chosen tetrahedron and is designated as the fixed worm tail, while the other appears on the neighbouring tetrahedron adjacent to the flipped spin and serves as the mobile worm head that subsequently propagates through the lattice. If the chosen tetrahedron already hosts a monopole of charge $\partial S$, the worm is instead initiated by annihilating this monopole and creating an identical monopole on a neighbouring tetrahedron. This process can be viewed as an {\em off-lattice entry} of the worm, with a fictitious (off-lattice) worm tail and the monopole on the neighbouring site acting as the mobile worm head~\cite{Kundu_Damle_2025_arxiv}. If the chosen tetrahedron carries any other charge, the worm attempt is rejected.

Once created, the worm head propagates via a weighted random walk on the dual diamond lattice, implemented through a sequence of spin flips that satisfy local detailed balance. At any stage of the propagation, the worm head encounters one of the following three situations.

(i) If the worm head arrives at a tetrahedron hosting a pre-existing monopole of equal and opposite charge, the two annihilate, and the worm terminates.

(ii) If the worm head encounters a pre-existing monopole of any other charge, it simply passes through without modifying that monopole.

(iii) In all remaining cases, where the worm head is located on a tetrahedron that was originally monopole-free, the worm head may either move to a neighboring tetrahedron or terminate via an {\em off-lattice exit}, which adds a pair of monopoles to the system.

Since off-lattice entry and exit respectively remove and add a pair of monopoles, the acceptance probabilities for these processes depend on the monopole fugacity and are chosen to ensure global detailed balance~\cite{Kundu_Damle_2025_arxiv}.

An important restriction of the algorithm is that a worm of species $\pm\partial S$ only modifies (the number and the positions of) monopoles carrying the same charge magnitude; monopoles of other charges remain unaffected. In particular, higher-charge worms {\em do not} branch into multiple lower-charge defects and therefore propagate on a restricted, depleted sublattice of flippable spins. For example, $\partial S=\pm3$ worms move only along dual diamond links hosting $S=\pm3/2$ spins. Despite this restriction, such worms substantially enhance equilibration deep in the $\ed<0$ regime, where $\pm3$ monopoles are deconfined.

Finally, unlike conventional algorithms that sample the ice manifold, worm propagation in this scheme may terminate at arbitrary length, allowing the creation or annihilation of charge pairs with a probability set by their fugacity. As a result, the algorithm operates in a grand-canonical ensemble for monopoles, with their equilibrium density controlled by temperature. In practice, we have used a Monte Carlo scheme in which each Monte Carlo step consists of a combination of local Metropolis single–spin updates together with non-local worm updates of species $\partial S=\pm1$ and $\partial S=\pm3$. The relative frequencies of each of these updates are chosen such that each produces $\mathcal{O}(N_s)$ spin flips per Monte Carlo step.

\section{II. Numerical Methods for quench dynamics}

To study the quench dynamics, we employ the waiting-time Monte Carlo (WTMC) algorithm~\cite{Dall_2001}. WTMC is a rejection-free local update scheme that enables access to long physical timescales --unattainable using conventional Glauber dynamics -- by explicitly associating a Monte Carlo time increment with each accepted update. In particular, even when a proposed local spin flip carries a large energy cost and hence an exponentially small acceptance probability, the move is still accepted, with the Monte Carlo time incremented by an amount inversely proportional to that probability. This feature allows the algorithm to efficiently capture the ultraslow relaxation dynamics of the $\delta$ excitations.

The quench protocol proceeds as follows. We first generate an equilibrium thermal ensemble at a high initial temperature $T_i$ for the model with $\jzz=1.0$, considering several values of $\Delta$ chosen symmetrically around $\Delta = 2\jzz$. Starting from a spin configuration drawn from this ensemble, the system is instantaneously quenched to a lower temperature $T_q$ and subsequently evolved using the local spin-flip WTMC dynamics. During the time evolution, we measure the densities of monopole excitations, $\rho_{|\mathcal{Q}|}$, and $\delta$ excitations, $\rho_{\delta}$, as functions of Monte Carlo time. All reported results are obtained by averaging over multiple independent realizations of the dynamics, each initialized from a different equilibrium configuration drawn from the high-temperature ensemble.

\end{document}